\documentclass[superscriptaddress,nofootinbib,twocolumn]{revtex4}
\usepackage{amsmath,lscape,epsfig}
\usepackage{amsfonts}
\usepackage{amsmath}
\usepackage{amssymb}
\usepackage{slashed}
\usepackage{hyperref}
\usepackage{color,soul}
\usepackage[utf8]{inputenc}
\usepackage{graphicx}
\usepackage{epstopdf}
\usepackage{ulem}

\def\ni{\noindent}
\def\beq{\begin{equation}}
\def\eeq{\end{equation}}
\def\beqa{\begin{eqnarray}}
\def\eeqa{\end{eqnarray}}
\def\ban{\begin{eqnarray*}}
\def\ean{\end{eqnarray*}}
\def\bi{\begin{itemize}}
\def\ei{\end{itemize}}

\begin{document}

\title{ Neutral meson properties in hot and magnetized quark matter: a new magnetic field 
independent regularization scheme applied to NJL-type model}

\author{Sidney S. Avancini} \email{sidney.avancini@ufsc.br}
\affiliation{Departamento de F\'{\i}sica, Universidade Federal de Santa
  Catarina, 88040-900 Florian\'{o}polis, Santa Catarina, Brazil}

\author{Ricardo L. S. Farias}
\email{ricardo.farias@ufsm.br}
\affiliation{Departamento de F\'{\i}sica, Universidade Federal de Santa Maria,
97105-900 Santa Maria, RS, Brazil}

\author{William R. Tavares} \email{williamr.tavares@hotmail.br}
\affiliation{Departamento de F\'{\i}sica, Universidade Federal de Santa
  Catarina, 88040-900 Florian\'{o}polis, Santa Catarina, Brazil}  

\begin{abstract}

 A magnetic field independent regularization scheme (zMFIR) based on the Hurwitz-Riemann
 zeta function is 
 introduced. 
  The new technique is applied to the regularization of the mean-field
  thermodynamic potential and mass gap equation within
  the SU(2) Nambu-Jona-Lasinio model in a hot and magnetized medium. The equivalence of the
  new and the standard
   MFIR scheme is demonstrated. The neutral meson pole mass is calculated in a hot and 
   magnetized medium and the advantages of using
  the new regularization scheme are shown.
\end{abstract}

\maketitle
\vspace{0.50cm}

\section{Introduction}
 
The possibility of strong magnetic fields of the order of $\sim 10^{19}$ G~\cite{fukushima01} 
or larger to be generated in non-central heavy-ion collisions has been a subject of great interest in the 
last decades, opening the possibility for new and interesting physical phenomena. The time 
scale and magnitude of such fields have been estimated in simulations
~\cite{skokov,skokov02,voronyuk,lou} for energies and impact parameters accessible in collisions 
of the Large Hadron Collider(LHC) and Relativistic Heavy Ion Collider (RHIC) . It is also expected that in
magnetars\cite{duncan,kae}, i. e., neutron stars with ultra-strong magnetic fields, 
magnetic fields  with magnitude  as large as $\sim 10^{18}$ G can be found inside the star.
In this way, the study of nuclear matter properties in a magnetized medium has attracted
the attention of many researchers nowadays. Usually these properties are calculated within the framework of
effective theories or lattice, once the non-perturbative behavior of quantum chromodynamics (QCD) 
prevents first principle evaluations at the low energy regime. Some current research problems of 
interest are the properties associated with the charged and neutral pion decay constants\cite{bali01,simonov03,iranianos}, decay width of vector modes\cite{sarkar03,band} and the masses of heavy mesons\cite{aguirre02,tetsuya,dudal04,kevin,gubler, noronha01,morita,morita02}. 
The masses of soft mesons have been studied in several approaches in effective models
\cite{nosso1,nosso03,zhuang,iran,scoccola01,huang01,scoccola02,luch,farias01,mao01, sarkar,sarkar02, zhang01,huan02,simonov01,fraga01,
aguirre,taya,shinya,andersen01,kojo,simonov02}, holografic QCD models\cite{dudal,dudal02} as well as 
QCD lattice simulations\cite{luschev,luschv02,luschv03,bali02,bali03,hidaka}. There are also some works involving light baryons\cite{andrei02,he}. 
 
In some recent papers~\cite{norberto,ricardo} the importance of using an appropriate regularization 
scheme to describe magnetized quark matter has been clearly demonstrated. In particular, it is shown in these references a strong dependence 
on the choice of the regularization scheme for the calculation of physical observables and more
importantly 
that an inappropriate regularization scheme may give rise to spurious solutions. For example, for the 
calculation of the magnetization some authors find oscillations which are unphysical, 
others find imaginary meson 
masses~\cite{iranianos} which are in fact spurious solutions due to the inappropriate choice of the 
regularization. This can be seen by comparing the meson masses results of reference~\cite{nosso1} where 
only real masses are obtained using  an adequate regularization scheme.  
The problem comes out when B-dependent regularization schemes are used. Therefore, it is
important to obtain regularization schemes where the separation between magnetic and non-magnetic
effects are made in a clean way, these schemes were baptized MFIR (magnetic field independent 
regularization) in ref~\cite{norberto}. As discussed in Ref.~\cite{ricardo}, some types of regularization prescriptions that do not separate exactly the contribution of the vacuum from the thermodynamic potential as, for instance, the Woods-Saxon or Lorentzian method, can result in unphysical oscillations in quantities such
as the effective quark masses or diquark condensates.
In ref.~\cite{klimenko} the Schwinger proper time method was used in order to regularize the NJL model
in a 
MFIR scheme, in refs.{~\cite{nosso1,nosso2} dimensional regularization was used. Both calculations 
obtaining completely equivalent results. In fact, this suggests that the separation in magnetic and 
non-magnetic
contributions is unique.
Although these techniques are extremely useful, there are situations where it is not simple or
inviable to use them. For example, the calculation of meson pole masses in a magnetized medium 
involves the calculation of  mesonic polarization loops with poles, which is not 
simple using the standard MFIR regularization of refs.~\cite{klimenko,nosso2}. Therefore, 
we are going to introduce next an alternative
MFIR scheme which is more general and may be used in several situations where the previous schemes 
are not applicable and in certain cases with clear advantages. Our objective is to treat 
effective models 
under strong magnetic fields like NJL models, 
linear and non-linear models, etc. To accomplish our goal we adapt the formalism of ref.~\cite{DE} 
where 
a rather comprehensive study of a magnetized relativistic electron gas is given in terms of the
Hurwitz-Riemann zeta function. In the last paper  the grand canonical potential, the magnetization,
the density, etc, are given analytically in terms of the Hurwitz-Riemann zeta function in a
very elegant way.  In this paper a new regularization 
scheme based on the Hurwitz-Riemann 
zeta function $\rm{z}$MFIR is introduced. In several situations there are advantages of this new technique 
compared to the usual MFIR.  From the numerical point of view the equations are easier to handle and thus opening up the
possibility of more complicated numerical calculations. For instance, the calculation
of susceptibilities can be done in a much more efficient way by using zMFIR since
we are dealing with several Hurwitz zeta functions instead of more
complicated functions.
We can use this formalism to obtain several  thermodynamical quantities
(e. g.  the magnetization, sound velocity, the specific heat and so on) 
through the calculation of derivatives which are easier 
to handle with the new formalism. Also, the analytical structure of the formalism
presented here
give us a more transparent way to see individually the importance of
the contributions from 
the medium and from the external magnetic field,
since the separation of the magnetic from
non-magnetic contributions are done in a exact way. In this way, we can also explore certain limits,
as well as asymptotic expansions.
 In the study of Bose-Einstein condensation in a magnetic field a pertinent formalism using the Hurwitz-Riemann zeta functions has been applied~\cite{bofeng1,bofeng2}.

The work is organized as follows. In Sec.~\ref{zmfir} we introduce the zMFIR formalism and apply in the
regularization of the mean-field
  thermodynamic potential and quark mass gap equation within
  the SU(2) Nambu-Jona-Lasinio model in a hot and magnetized medium. In Sec.~\ref{pi0mass} the neutral
  meson pole mass is calculated 
  in a hot and magnetized medium using zMFIR formalism. In Sec.~\ref{numresult} we present our numerical results and compare them with 
  the recent literature . In Sec.~\ref{imc} we introduce thermo-magnetic effects in the NJL coupling
  constant~\cite{farias,farias2} to
  include effects due the inverse magnetic catalysis effects on the meson masses at finite temperature 
  and magnetic fields. Finally, 
  in Sec.~\ref{conclusions} we discuss our results and conclude. We leave
  for the appendix the explicit calculations 
  of some important quantities.
%
\section{ $\rm{z}$MFIR - a regularization scheme based on the Hurwitz-Riemann zeta function}
\label{zmfir}
In order to motivate and adapt the formalism of the authors of ref.~\cite{DE} to be 
used in effective models of QCD we discuss in a rather different way the development of 
these authors in what follows. We are specially interested in the study of the 
thermodynamics of effective models. It is well known that all the thermodynamical properties
can be obtained from the partition function or the Grand canonical potential $\Omega$:
\begin{equation}
  \cal{Z} = \rm{Tr} ~e^{-\beta \left( \hat{H} - \mu \hat{N} \right) } ~,~\Omega = -\frac{1}{\beta}\log \cal{Z}
\end{equation}
where $\hat{H}$ is the Hamiltonian operator of the system, $\beta$ is the inverse of
the temperature and $\mu$ is chemical potential. Here, we 
restrict our discussion to the case where there is only one chemical specie, the generalization 
for several species is trivial and for our line of reasoning it is an unnecessary and irrelevant 
complication. The partition function may be calculated through several techniques, like second 
quantization, 
Feynman functional integration, etc.
For the calculation of $\cal{Z}$ one of the approximations most used 
in the literature is the large N (or mean field approximation (MFA)). In a large class of 
models, the Grand canonical potential 
and the thermodynamical properties derived from it, can be written as:
\begin{equation}
 I=\sum_f \sum_{s=\pm 1} \int \frac{d^3 p}{(2\pi)^3} f(E_f)~~,~~ E_f=\sqrt{p^2+M_f^2}~~,
\end{equation}
where the summation is realized over the flavor and the spin projection in 
direction-3, $f(E_f)$ is a function of the energy, $E_f$,
and often the integral $I$ contains divergent parts that have to be regularized by an 
appropriate method, like
Pauli-Villars, sharp covariant or non-covariant cutoff, form factors, etc.
Our aim here is to consider fermionic physical systems consisting of a charged particle immersed 
in a hot and magnetized medium. Assuming without any loss of 
generality the magnetic field $B$ in the  z-direction,   it is well 
known~\cite{revfraga,revandersen,revigor} that the motion is  quantized in the plane perpendicular to the z-axis 
resulting  in a sum over Landau levels and  a dimensional reduction. From an heuristic point of view 
one can pass from non-magnetized to magnetized systems through the prescription:
\begin{eqnarray}
 I_f(0)&=&  I_f(B=0) = \sum_{s=\pm 1} \int \frac{d^3 p}{(2\pi)^3} f(E_f)~~   \nonumber \\ 
 \rightarrow &&
 I_f(B)=\beta_f \sum_{n=0}^\infty g_n \int_{-\infty}^{\infty}
 \frac{dp_3}{(2\pi)^2}f(E_n)~~,~~ \label{IB}    \\
  E_f&=&\sqrt{\vec{p}^{~2}+M_f^2}~~\rightarrow  ~~E_n=\sqrt{p_3^2+M_f^2+2\beta_f n}~, \nonumber
\end{eqnarray}
where $s=\pm 1$, $n=0,1,2,\dots$ stand for the spin and the Landau 
levels respectively, $g_n=2-\delta_{n0}$ is the degeneracy factor, 
$\beta_f=|q_f|B$ with $|q_f|$ the absolute value of the electric charge of the particle $f$ and $M_f$ 
its mass. In particular, in the present work we consider only situations where 
the latter prescription can be used.
Here, we introduce the key ingredient that will allow us to formulate the new 
regularization scheme, i. e., 
the non-normalized density of states, $ g_f(E,B)$,
\begin{equation}
  g_f(E,B)=\frac{\beta_f}{(2\pi)^2}\sum_{n=0}^\infty g_n \int_{-\infty}^{\infty}
  dp_3 ~\delta(E-E_n)~~.\label{gE}
\end{equation}
 The main idea is to use integrals involving the density of states instead of
momentum integrals.  
After substituting the last expression in following integral:
\begin{equation}
 I_f(B)=\int_{M_f}^{\infty} dE~g_f(E,B)~f(E)~, \label{IMFIR}
\end{equation}
one straightforwardly recovers  $I_f(B)$ from the eq.(\ref{IB}).
Next, using the following property of the Dirac delta function:
\begin{equation}
 \int_{-\infty}^\infty~dx~\delta(h(x))~=~\sum_i ~\frac{1}{|h^\prime (x_i)|}~,
\end{equation}
where $x_i$ are the roots of $h(x)$, i.e., $h(x_i)=0$, $i=1,2,\dots$ and $h^\prime(x)=\frac{dh(x)}{dx}$,
we are able to perform in eq.(\ref{gE}) the  integral in $p_3$ resulting in:
\begin{equation}
  g_f(E,B)=\frac{2\beta_f}{(2\pi)^2}\sum_{n=0}^{n_{max}} g_n 
  \frac{E}{(E^2-M_f^2-2\beta_f~n)^{1/2}}~~,\label{gE2}
\end{equation}
where $n_{max}=[\frac{E^2-M_f^2}{2\beta_f}]$ with $[x]$ being the floor function, i. e., the function 
which gives the largest integer less than or equal $x$. Defining 
, $q_E \equiv \frac{E^2-M_f^2}{2\beta_f}$, we rewrite 
eq.(\ref{gE2}) as:
\begin{equation}
  g_f(E,B)=\frac{(2\beta_f)^{1/2}}{(2\pi)^2}~E \sum_{n=0}^{[q_E]} g_n \frac{1}{(q_E-n)^{1/2}}
  ~~.\label{gE3}
\end{equation}
In order to obtain a convenient representation of $g_f(E,B)$ we substitute  in the summation 
of eq.(\ref{gE3})
the explicit form of the degeneracy factor $g_n$ yielding:
\begin{equation}
\sum_{n=0}^{[q_E]} g_n \frac{1}{(q_E-n)^{1/2}} = \sum_{n=0}^{[q_E]} 
\frac{2}{(q_E-n)^{1/2}} -\frac{1}{q_E^{1/2}} ~~ .
\end{equation}
The last sum may be done using the Hurwitz-Riemann zeta function~\cite{apostol,DE}
\begin{equation}
  \zeta(z,q)=\sum_{k=0}^{\infty}\frac{1}{(q+k)^{z}} \label{HZ}
\end{equation}
and its property:
\begin{equation}
 \sum_{n=0}^{N}\frac{1}{(q+n)^z}=\zeta(z,q) -\zeta(z,q+N+1)
\end{equation}
Thus, after some straightforward manipulations one obtains the following expression~\cite{DE}
for $q_f(E,B)$:
\begin{eqnarray}
&& g_f(E,B)=E~\frac{(2\beta_f)^{1/2}}{(2\pi)^2}~\nonumber \\
 \times  &&\left[ 2\left( \zeta(\frac{1}{2},\{q_E\})-\zeta(\frac{1}{2},q_E+1)\right) -\frac{1}{q^{1/2}_E} \right] ~~,
 \label{gEZ}
\end{eqnarray}
where $\{ q_E\} \equiv q_E-[q_E]$ is the fractional part of $q_E$, the last expression can be written
in a simplified way using the Hurwitz-Riemann zeta 
function property~\cite{apostol}:
\begin{equation}
 \zeta(z,q+1)=\zeta(z,q)-\frac{1}{q^z} ~~, \nonumber
\end{equation}
resulting in:
\begin{eqnarray}
&& g_f(E,B)=E~\frac{(2\beta_f)^{1/2}}{2\pi^2}~\nonumber \\
 \times  &&\left( \zeta(\frac{1}{2},\{q_E\})-\zeta(\frac{1}{2},q_E) +
 \frac{1}{2q^{1/2}_E} \right) ~~.
 \label{gEZ2}
\end{eqnarray}

The last expression do not involve anymore
the explicit sum over Landau levels, however, the magnetic and non-magnetic contributions are
still entangled.
In order to separate the magnetic from the non-magnetic contribution 
we take the limit  $(B \rightarrow 0)$ or equivalently
$(\beta_f \rightarrow 0$ and $q_E \rightarrow \infty)$ in eq.(\ref{gEZ2}). This limit can be calculated  
using the Hurwitz-Riemann zeta function asymptotic limit~\cite{DE} 
\begin{equation}
\zeta(\frac{1}{2},q_E)=-2q_E^{1/2}+\frac{1}{2}q_E^{-1/2}+\frac{1}{24}q_E^{-3/2}+O(q_E^{-7/2})~~,
\label{asym}
\end{equation}
and noting that $\zeta(1/2,\{q_E\})$
is a periodic and limited function of $q_E$,
therefore, one obtains:
\begin{equation}
 \lim_{\beta_f \to 0} g_f(E,B)=E~\frac{(2\beta_f)^{1/2}}{2\pi^2}~2q_E^{1/2}
 =\frac{E\sqrt{E^2-M_f^2}}{\pi^2}\nonumber \\
 ~.
\end{equation}
In the latter expression  the density of states of the non-magnetized system is recovered. 
Of course, this result 
should be expected since taking the limit ${B \to 0}$ in the density of states one 
has to recover the $B=0$ expression. At this point we write for future convenience:
\begin{equation}
 g_f(E,B)~=~g_f(E)~+~ \bar{g}_f (E,B) ~~ , \label{gzMFIR}
\end{equation}
where $g_f(E)$ is the non-magnetic contribution,
\begin{equation}
 g_f(E) = \frac{E\sqrt{E^2-M_f^2}}{\pi^2} ~~ \label{nonmag}
\end{equation}
and $\bar{g}_f(E,B)$ is the corresponding purely magnetic contribution:
\begin{eqnarray}
 \bar{g}_f (E,B) &\equiv& \left( g_f(E,B)-E~\frac{(2\beta_f)^{1/2}}{2\pi^2}~2q_E^{1/2}
 \right) \nonumber \\
 &=& E~\frac{(2\beta_f)^{1/2}}{2\pi^2}~\nonumber \\
\times \left( \zeta(\frac{1}{2},\{q_E\}) \right. &-& 
 \left. \zeta(\frac{1}{2},q_E) -2q_E^{1/2}+\frac{1}{2q_E^{1/2}} \right) ~~. 
\end{eqnarray}

The density of states written in the form of eq.(\ref{gzMFIR}) separates in an exact 
way the magnetic from the 
non-magnetic contribution. Therefore,  to calculate any physical 
expression, we use $g_f(E,B)$ together with eq.(\ref{IMFIR}). Of course, divergences
may  still be 
present and have 
to be regularized using any appropriate regularization scheme.  
From eq.(\ref{asym}) one easily notices that
$\bar{g}_f(E,B)$ goes to zero 
when $B \rightarrow 0$ as expected.
In conclusion, we have achieved through eq.(\ref{gzMFIR}) our objective,i. e., to separate
in the calculation 
of any physical expression the magnetic from 
the non-magnetic contribution. We name the present  approach (zMFIR), i. e., 
a zeta function based magnetic field independent regularization  scheme. 
For the practical use of eq.(\ref{IMFIR})
we write:
\begin{equation}
 I_f(B)=I_f (0) +\tilde{I}_f (B)~~, \label{intzMFIR}
\end{equation}
where from eq.(\ref{nonmag}) the non-magnetic contribution is given in terms of the 
momentum variable, $p$ = $\sqrt{E^2-M_f^2}$, by:
\begin{equation}
  I_f(0)=\int_{M_f}^{\infty} dE~g_f(E)~f(E_f)~=~ \sum_{s=\pm 1} 
  \int \frac{d^3p}{(2\pi)^3}~f(E_f)~.
  \label{intB0}
\end{equation}
Analogously, the magnetic contribution can be obtained in  terms of the variable $q_E$ as:
\begin{eqnarray}
  \tilde{I}_f(B)&=&\int_{M_f}^{\infty} dE~\bar{g}_f(E,B)~f(E) \nonumber \\
  ~&=&~ \int_{0}^{\infty} ~dq_E~ \tilde{g}_f(q_E,B)~f(E(q_E))~, \label{intB}
\end{eqnarray}
with
\begin{equation}
 E(q_E) = \sqrt{M_f^2+2\beta_f~q_E} \label{EqE} ~~,
\end{equation}
and the corresponding density of states as a function of $q_E$:
\begin{equation}
\tilde{g}_f(q_E,B) = ~\frac{(2\beta_f)^{3/2}}{(2\pi)^2}\times \tilde{\cal H}_{1/2}(q_E) 
\label{gEZETA}~,
\end{equation}
where
\begin{equation}
\tilde{\cal H}_{1/2}(q_E) = \left( \zeta(\frac{1}{2},\{q_E\})  - 
  \zeta(\frac{1}{2},q_E) -2q_E^{1/2}+\frac{1}{2q_E^{1/2}} \right) ~~.
 \label{gEZETA1}  
\end{equation}
Some comments are in order, with the present formalism we have summed over 
the Landau levels and 
obtained expressions which are integrals 
in terms of zeta functions.
When considering the numerical calculation of such integrals some care has to be 
done since the 
$\zeta(\frac{1}{2},\{q_E\})$ is a periodic function of $q_E$ with period 1. In 
some cases, like in the 
calculation of the NJL thermodynamic potential the periodic function can be
treated in very efficient way, 
however, in general, there are efficient numerical algorithms to perform quadrature 
of periodic 
functions~\cite{intbook}.
In the next section we consider the regularization of the two flavor NJL  mean-field 
thermodynamic potential
using the zMFIR formalism.
%
%
%
%
\subsection{The su(2)-NJL regularized thermodynamic potential}
\noindent  The NJL model lagrangian~\cite{NJL-122,buballa,kleva} is given by:
\begin{eqnarray}
\mathcal{L}&=&\overline{\psi}\left(i \slashed D - \tilde{m}\right)\psi
+G\left[(\overline{\psi}\psi)^{2}+(\overline{\psi}i\gamma_{5}\vec{\tau}\psi)^{2}\right]\nonumber \\
&-&
\frac{1}{4}F^{\mu\nu}F_{\mu\nu} ~,
\end{eqnarray}
where $A^\mu$ is the electromagnetic gauge field, 
$F^{\mu\nu} = \partial^\mu A^\nu - \partial^\nu A^\mu$ , $\vec{\tau}$ is the isospin matrix,  G  is 
the coupling constant,
Q=diag($q_u$= $2 e/3$, $q_d$=-$e/3$) is the charge matrix,
$D^\mu =(i\partial^{\mu}-QA^{\mu})$ is the covariant derivative,   
 $\psi$ is the quark fermion field and $\tilde{m}$ represents the bare quark mass matrix,
\begin{equation}
 \psi = \left(
\begin{array}{c}
\psi_u  \\
\psi_d \\
\end{array} \right) ~, ~
 \tilde{m}  = \left(
\begin{array}{cc}
m_u & 0 \\
0 & m_d \\
\end{array} \right) ~,
\end{equation}
where we take $m$=$m_u$=$m_d$ and adopt the Landau gauge, i. e., $A^{\mu}=\delta_{\mu 2}x_{1}B$, 
thus $\vec{B}=B{\hat{e_{3}}}$. In the mean field approximation
 the NJL Lagrangian is given by~\cite{buballa}: 
\begin{equation}
 \mathcal{L}=\overline{\psi}\left(i\slashed D-M\right)\psi+G \left \langle \overline{\psi}\psi \right \rangle^{2}-
 \frac{1}{4}F^{\mu\nu}F_{\mu\nu}~,
\end{equation}
\noindent
\begin{equation}
 M=m-2G \left \langle \overline{\psi}\psi \right \rangle. \label{gap}
\end{equation}
 where $\langle \overline{\psi}\psi  \rangle $ is the quark condensate. The mean-field 
 thermodynamic potential with $B=0$ is given~\cite{buballa,kleva}:
\begin{equation}
 \Omega(T,\mu,M) =\Omega_{vac} +\Omega_{\cal M}(T,\mu) ~+ ~\frac{(M-m)^2}{4G}  + 
 {\rm const.} ~, \label{PTB0}
\end{equation}
 with
\begin{equation}
  \Omega_{vac}=
   -N_c \sum_{f=u,d} \sum_{s=\pm 1}  \int \frac{d^3p}{(2\pi)^3} E  ~.   \label{Evac}
 \end{equation}
 \begin{equation}
 \Omega_{\cal M}(T,\mu)=-2N_fN_c \int \frac{d^3p}{(2\pi)^3} \left[  T\ln \left( 1+ e^{-\beta (E-\mu)} \right)
                                                                 \right. \nonumber
 \end{equation}
 \begin{equation}
 ~+~  \left.
 T\ln \left( 1+ e^{-\beta (E+\mu)} \right)
 \right] ~,
 \end{equation}
\ni  with $E=\sqrt{p^2+ M^2}$, $N_f$=2 and $N_c$=3. In a magnetized medium the corresponding
mean-field 
thermodynamic potential can be obtained through the prescription, Eq.(\ref{IB}):

 \begin{equation}
  \Omega_{vac}=- N_c \sum_{f=u,d} \beta_f \sum_{n=0}^\infty  g_n \int_{-\infty}^{\infty}
 \frac{dp_3}{(2\pi)^2} E_n ~, \label{potmag}
 \end{equation}
 \begin{eqnarray}
 && \Omega_{\cal M}(T,\mu)=-N_c \sum_{f=u,d}\beta_f \sum_{n=0}^\infty  g_n  
 \int_{-\infty}^{\infty}
 \frac{dp_3}{(2\pi)^2}  \\ \nonumber
 &&\times \left[  T\ln \left( 1+ e^{-\beta (E_n-\mu)} \right)
 ~+~  
 T\ln \left( 1+ e^{-\beta (E_n+\mu)} \right)
 \right] ~, \label{Emed}
 \end{eqnarray}
where $E_n=\sqrt{p_3^2+M_f^2+2\beta_f n}$ with $\beta_f=|q_f|B$, $f$=$u,d$.
This latter expression has also been derived in an alternative way in ref.~\cite{nosso2}. 
From eqs.(\ref{intzMFIR},\ref{intB0},\ref{intB}) the thermodynamic potential in a 
magnetized medium can be written as:
\begin{equation}
   \Omega(T,\mu,M,B) = \Omega(T,\mu,M) +\tilde{\Omega}_{vac}(B) +
   \tilde{\Omega}_{\cal M}(T,\mu,B) ~,
\end{equation}
where $\Omega(T,\mu,M)$, i. e., the $B=0$ term is given in eq.(\ref{PTB0}) and  
 \begin{equation}
  \tilde{\Omega}_{vac}(B)=-N_c~ \sum_{f=u,d} \int_{0}^{\infty} ~dq_E~ \tilde{g}_f(q_E,B)~E(q_E) ~,
  \label{EvacB}
 \end{equation}
 \begin{eqnarray}
 && \tilde{\Omega}_{\cal M}(T,\mu,B)=-N_c~T \sum_{f=u,d} \int_{0}^{\infty} ~dq_E~
 \tilde{g}_f(q_E,B)~\\ \nonumber
 && \times \left[ \ln \left( 1+ e^{-(E(q_E)-\mu)/T} \right)
 +  
 \ln \left( 1+ e^{-(E(q_E)+\mu)/T} \right)
 \right] ~. \label{EmedB}  
 \end{eqnarray}
Now, we proceed with the zMFIR regularization scheme, the only divergent terms of
the thermodynamic potential are  
 the vacuum terms given by eq.(\ref{Evac}) and eq.(\ref{EvacB}). 
The term $\Omega_{vac}$ is the usual ultraviolet divergent NJL vacuum and we use
a 3D non-covariant cutoff $\Lambda$ for its regularization. The magnetic vacuum 
$\tilde{\Omega}_{vac}(B)$ is also ultraviolet divergent and, hence, 
we discuss its regularization
in Appendix~\ref{appA}, where it is shown that the finite magnetic vacuum contribution is:
 \begin{eqnarray}
&&\tilde{\Omega}_{vac}(B)=-N_c\sum_{f=u,d} \frac{(2\beta_f)^{2}}{(2\pi)^2} 
                                         ~ \left[ \frac{1}{24} \ln x_f + \frac{2}{24} \right.
                                         \nonumber \\ 
&& -\int_0^{1}dq_E ~ \zeta(-\frac{1}{2},q_E) ~\zeta(\frac{1}{2},x_f+ q_E) 
                                                     \nonumber \\
&& +\int_0^{\infty}dq_E \left( ~ \zeta(-\frac{1}{2},q_E) + \frac{2}{3}q_E^{3/2} \right.
                                                     \nonumber \\
&&\left. \left.  -\frac{1}{2}q_E^{1/2} +\frac{1}{24}q_E^{-1/2} \right)
                                              \frac{1}{\sqrt{x_f+q_E}}   \right]    ~, \label{vaczeta}
 \end{eqnarray}
where $x_f=\frac{M_f^2}{2\beta_f}$. This last magnetic vacuum term was
also calculated using MFIR and dimensional 
regularization in Ref.~\cite{nosso2} 
where the following expression was obtained:
 \begin{eqnarray}
&&\tilde{\Omega}_{vac}(B)=-N_c\sum_{f=u,d} \frac{(\beta_f)^{2}}{2\pi^2} 
                                                 ~ \left[  \zeta^\prime(-1,x_f) \right. \nonumber \\ 
                                                     \nonumber \\
&&\left.   -\frac{1}{2} [x_f^2-x_f ]\ln x_f +\frac{x_f^2}{4}  \right]  ~. \label{vacdr}
 \end{eqnarray}
Of course, in this particular situation the latter expression is much simpler for numerical calculations. 
Nevertheless, in table-I are shown the FORTRAN numerical results using both expressions and the
respective relative error. A constant term which drops out when physical observables~\cite{nosso2}
are calculated was added for the comparison. It is clear from the numerical calculation the equivalence between the two methods.
\begin{table}[ht]
\caption{Comparison between the vacuum results obtained according to eq.(\ref{vacdr}), $E_{vac1}$,
and eq.(\ref{vaczeta}), $E_{vac2}$ and their relative errors,
$Error$= $|E_{vac1}$-$E_{vac2}|$/$E_{vac1}$ for the su(2)-NJL model at $eB = 0.1 GeV^2$ and several
effective masses.}
\begin{center}
\begin{tabular}{cccc}
\hline
$M_{eff}$      &  $E_{vac1}$ (GeV)  & $E_{vac2}$ (GeV) &   Error 
\\ \hline
  408.93     & -0.9664340125$\times 10^{-4}$  &  -0.9664338550$\times 10^{-4}$ &  1.63$\times 10^{-7}$   \\
  368.82     & -0.8239054256$\times 10^{-4}$  &  -0.8239052680$\times10^{-4}$  &  1.91$\times 10^{-7}$   \\
  118.43     &  0.5487198936$\times 10^{-4}$  &   0.5487200511$\times 10^{-4}$ &  2.87$\times 10^{-7}$   \\
   43.19     &  0.1163457449$\times 10^{-3}$  &   0.1163457607$\times 10^{-3}$ &  1.35$\times 10^{-7}$   \\ 
   10.95     &  0.1369580943$\times 10^{-3}$  &   0.1369581101$\times 10^{-3}$ &  1.15$\times 10^{-7}$  \\
\hline
\end{tabular}
\end{center}
\label{tab1}
\end{table}
\noindent
The mass gap expression for the non-magnetized NJL model is given by~\cite{buballa,kleva}:
\begin{equation}
 \frac{M-m}{2MG}=I_G+I_G(T,\mu)  ~.
\end{equation}
\ni with
\begin{eqnarray}
&& I_G = 2 N_c N_f \int \frac{d^3 p}{(2\pi)^3} \frac{1}{\sqrt{p^2+ M^2}}  \label{gapB0} \\
&& I_G(T,\mu) = -2 N_c N_f \int \frac{d^3 p}{(2\pi)^3} 
         \frac{n(E)+\bar{n}(E)}{\sqrt{p^2+ M^2}}  \nonumber
                                  \label{gapB0T} ~, \\
\end{eqnarray}
with the Fermi distribution functions of particles, $n(E)$, and anti-particles, $\bar{n}(E)$,
given by:
\begin{equation}
 n(E)=\frac{1}{1+e^{(E-\mu)/T}} ~~,~~ \bar{n}(E)=\frac{1}{1+e^{(E+\mu)/T}} ~.
\end{equation}
\noindent
The magnetized gap equation was obtained in Ref.~\cite{nosso1,nosso2} and is given by:
\begin{multline}
 \frac{M-m}{2MG}=
N_c\sum_{f=u,d}~\beta_f \sum_{n=0}^{\infty} g_{n} \\  
 \times\int_{-\infty}^{\infty} \frac{dp_3}{(2\pi)^2} 
 \frac{ 1}{ \sqrt{p_{3}^{2}+M^{2}+2\beta_{f} n} }  (1-n(E_n)-\bar{n}(E_n)) 
  \\  ~.             \label{gap_eq}
\end{multline}
Note that the latter expression may be obtained directly from eqs.(\ref{gapB0},\ref{gapB0T}) using the
prescription given in eq.(\ref{IB}). Thus, from the zMFIR 
scheme, eqs.(\ref{intzMFIR},\ref{intB0},\ref{intB}),
one obtains:
\begin{equation}
 \frac{M-m}{2MG}=~I_G+I_G(T,\mu)+I_G(B)+I_G(T,\mu,B)~,\label{gap_zmfir}
\end{equation}
where the vacuum term, eq.(\ref{gapB0}), is the only divergent term which we regularize through a
non-covariant 3D cutoff obtaining as usual: 
\begin{equation}
I_G= \frac{N_c}{\pi^2} 
\left( \Lambda \epsilon_\Lambda -M^2\ln\left( \frac{\Lambda+\epsilon_\Lambda }{M} \right)
\right)  \nonumber  ~,
\end{equation}
where $\epsilon_\Lambda= \sqrt{\Lambda^2+M^2}$ and $I_G(T,\mu)$ is the finite non-magnetic temperature 
(or medium) dependent term given in eq.(\ref{gapB0T}). The finite magnetic contributions are given by:
%
%
\begin{eqnarray}
&& I_G(B)=N_c \sum_{f=u,d}~\sum_{n=0}^{\infty}  g_{n}\beta_{f} 
 \int_{-\infty}^{\infty} \frac{dp_3}{(2\pi)^2}  \frac{ 1}{E_n}  \nonumber \\
 && = N_c\sum_{f=u,d}\int_{0}^{\infty} dq_E \frac{(2\beta_f)^{3/2}}{(2\pi)^2}
 \times \tilde{\cal H}_{1/2}(q_E) 
                      \frac{1}{E(q_E)} ~, \label{gapzmfir}
\end{eqnarray}
\begin{eqnarray}
 I_G(B,T,\mu)&=&-N_c \sum_{f=u,d}~\sum_{n=0}^{\infty}  g_{n}\beta_{f}\nonumber \\
&\times& \int_{-\infty}^{\infty} \frac{dp_3}{(2\pi)^2}  \frac{ \left(n(E_n)+\overline{n}(E_n) \right) }{E_n}
   \nonumber \\
  &=& -N_c\sum_{f=u,d} \int_{0}^{\infty} dq_E~\frac{(2\beta_f)^{3/2}}{(2\pi)^2}\times \tilde{\cal H}_{1/2}(q_E) 
  \nonumber\\
  &\times&
                       \frac{\left(n(E(q_E))+\overline{n}(E(q_E)) \right) }{E(q_E)} ~.
\end{eqnarray}
Therefore, we have obtained an  exact separation between the magnetic and non-magnetic terms. To our
knowledge this complete separation for the gap and thermodynamic potential has not yet been obtained 
elsewhere in the literature. 
The term $I_G(B)$ has also been derived using dimensional regularization in Ref.~\cite{nosso2}
and we have 
checked numerically that the latter calculation and the present one coincides. 
Note that eq.(\ref{gapzmfir})  may also be obtained directly from the derivative of 
the thermodynamic potential 
with respect to the effective mass $M$, $\partial \Omega(T,\mu,M,B) /\partial M$=0 yielding 
after some simple manipulations, exactly the same expression. This demonstrates the consistency of the 
zMFIR scheme. In the Appendix \ref{appB} we show 
analytically the equivalence between the zMFIR and the 
MFIR formalisms for the gap equation.
%
%
\section{The $\pi^0$ pole mass in a magnetized medium}
\label{pi0mass}
 Next, we will give a brief discussion of the $\pi^0$ mass~\cite{NJL-122,kleva} 
 calculation within the context of the zMFIR. This is an example where the usual 
 dimensional regularization (MFIR) performed in ref.~\cite{nosso1} for the zero temperature and
 chemical potential case, present problems. In particular, when one considers the system in a 
 magnetized medium due to the 
 existence of a real pole in the polarization function the formalism for temperatures
 above the Mott temperature is trick.  In the next section, we hope to make clear 
 the advantage of using the new 
 regularization scheme in the present case, since using zMFIR the the analytical structure 
 of the polarization function will
 be clearly exposed and the manifest effect of dimensional reduction will be seen in a transparent way.
 We use the  RPA approximation, 
which  formally consists in summing the
geometric series diagrammatically represented in Fig. \ref{RPA}. The left hand side of the 
equality in  Fig. \ref{RPA} 
can be  calculated by representing  the quark-pion interaction with the following Lagrangian~\cite{kleva}:
\begin{equation}
 \mathcal{L}_{\pi qq}=i g_{\pi qq} \overline{\psi} \gamma_5 \vec{\tau}\cdot \vec{\pi} \psi~,
\end{equation}
\noindent where  $\vec{\pi}$ stands for the pion field while $g_{\pi qq}$ represents the coupling strength
between pions and quarks.
Both sides of the equality in  Fig. \ref{RPA} can be calculated using standard Feynman rules and 
 the quark (dressed), $S_q(k^2)$,  as well as the meson, $D_{{\pi}^0}(k^2)$,   
 propagators\noindent~\cite{gus,kuz}:
\begin{equation}
  D_{\pi^0}(k^2) = \frac{1}{k^2-m_{\pi^0}^2} ~, \label{meson}
\end{equation}
\begin{equation}
 S_{f}(x,x')=e^{i\Phi_f (x,x')}\sum_{n=0}^{\infty}{S}_{f,n}(x-x')~,~f=u,d~, \label{prop1}
\end{equation}
\noindent  
the above propagator is given by the product of a gauge dependent factor, $\Phi_f(x,x')$, 
called Schwinger phase, times a translational invariant term and its explicit expression 
can be found in~\cite{kuz}.
Selecting the quantum numbers associated to the neutral pion, $\pi^0$, and noting that in this case the 
phase factors cancel out,
one can show that the effective interaction is given by the relation:
\begin{equation}
  (ig_{\pi^{0} qq})^2~iD_{{\pi}^0}(k^2)=\frac{2iG}{1-2G\Pi_{ps}(k^{2})} ~, \label{BS}
\end{equation}
where the pseudo-scalar polarization loop reads:
\begin{eqnarray}
\frac{1}{i}\Pi_{ps}(k^{2})&=&-\sum_{f=u,d} \int \frac{d^{4}p}{(2\pi)^{4}}
Tr\left[i\gamma_{5}iS_f (p+\frac{k}{2})
i\gamma_{5}\right. \nonumber\\
&\times&\left.i S_f (p-\frac{k}{2})\right]~.  \label{loop-ps}
\end{eqnarray}
Proceeding analogously one obtains for the scalar channel, 
\begin{equation}
\frac{1}{i}\Pi_{s}(k^{2})=-\sum_{f=u,d} \int \frac{d^{4}p}{(2\pi)^{4}}
Tr[iS_f(p+\frac{k}{2})iS_f(p-\frac{k}{2})]~.
\label{loop-sc}
\end{equation}
From eq.(\ref{BS}) one can obtain the $\pi^0$ mass pole as:
\begin{equation}
 1-2G\Pi_{ps}(k^{2})|_{k^2=m^2_{\pi^0}}=0~. \label{polemass}
\end{equation}
\begin{figure}[h]
\begin{tabular}{ccc}
\end{tabular}
\end{figure}
\begin{figure}[h]
\begin{tabular}{ccc}
\includegraphics[width=7.cm]{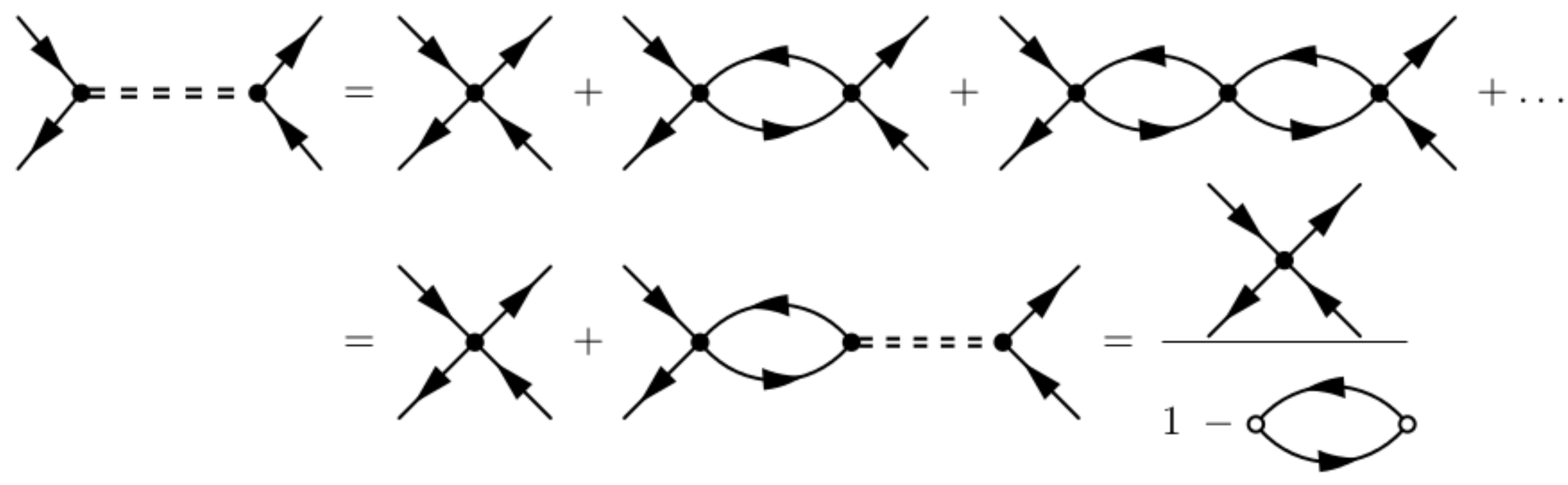}\\
\end{tabular}
\caption{Diagrammatic representation of the RPA approximation. }
\label{RPA}
\end{figure}

In Ref.~\cite{nosso1} the pseudo-scalar polarization loop and the $\pi_0$ pole mass 
have been calculated using the MFIR scheme. Next, we use the zMFIR regularization scheme
to rewrite the  expressions of the  latter reference for the 
calculation of the $\pi_0$ pole mass. We start from the 
non-magnetized expression for the $\pi_0$ pole mass~\cite{kleva}:
\begin{equation}
m_{{\pi}^0}^{2}=-\frac{m}{M}
\frac{1}{ 4G N_{c}N_{f} I(m_{\pi^0}^2) }~, \label{mpionB0}
\end{equation}
where in previous equation $k=(k_0=m_{\pi^0}^2,\vec{k}=\vec{0})$ and
\begin{equation}
I(k^2)=   
i\int \frac{d^{4}p}{(2\pi)^4}\frac{1}{[(p+\frac{k}{2})^{2}-M^{2}]
 [(p-\frac{k}{2})^{2})-M^2] }\, . \label{Iqn0}
\end{equation}
\ni
Here, after using the Matsubara formalism in the latter integral, one obtains:
\begin{equation}
I(k_0^2)=   
\int \frac{d^{3}p}{(2\pi)^3}\frac{1}{E(k_0^2-4E^2)}(1-n(E)-\bar{n}(E))\, , \label{IqnB0}
\end{equation}
where the integration with respect to the variable $p_0$ was done by using 
the residue method~\cite{PC1}. The magnetized version of the latter equation is obtained through
the use of the prescription given in eq.(\ref{IB}) and it can be written as a sum of two 
terms, i. e., the vacuum and medium contribution:
\begin{equation}
 I(k_0^2) \rightarrow I(k_0^2,B,T)~=~I_{vac}(k_0^2,B)+I_{T,\mu}(k_0^2,B)~, \label{polmed}
\end{equation}
where
\begin{eqnarray}
 I_{vac}(k_0^2,B) &=&
 \sum_{f=u,d} \beta_{f} \sum_{n=0}^{\infty}
  g_{n} \int_{-\infty}^{\infty}  \frac{d p_3}{(2\pi)^{2}}  \nonumber \\
&\times& \frac{1}{4 E_n\left(k_0^2-4E_n^2\right)}  ~, \label{mpion_2}
\end{eqnarray}
\begin{eqnarray}
I_{T,\mu}(k_0^2,B)&=&-
 \displaystyle{\sum_{f=u,d} \beta_{f} \sum_{n=0}^{\infty}
 g_{n}} \int_{-\infty}^{\infty}  \frac{d p_3}{(2\pi)^{2}}  \nonumber \\
 &\times& 
 \frac{1}{4 E_n\left(k_0^2-4E_n^2\right)} 
  \left(n(E_n)+\overline{n}(E_n) \right)~. \label{therm}
\end{eqnarray}
 The vacuum part of the previous expression was obtained in Ref.~\cite{nosso1} using the MFIR scheme.
 It was shown that it can be separated in two terms, one the usual $B=0$ vacuum  contribution 
 and another due to the pure magnetic contribution:
\begin{equation}
 I_{vac}(k_0^2,B)=I_{vac}(k_0^2)+I(k_0^2,B) ~,  \label{ivac4}
\end{equation}
\begin{equation}
 I_{vac}({k_0}^2) =  
\int  \frac{d^3p}{(2\pi)^3}\frac{1}{E\left(k_0^2-4E^2\right)} ~,  \label{ivac3}  
\end{equation}

%
%
\begin{eqnarray}
 I(k_0^2,B)&=&-\frac{\pi}{4(2\pi)^3}\sum_{f=u,d}
\int^{1}_{0} dx  \left[ -\psi\left( \bar{x}_f +1 \right)  +
\frac{1}{2\bar{x}_f}\right.\nonumber\\
& +&\left. \ln \bar{x}_f  \right]  ~, \label{int00} 
\end{eqnarray}
where  
\begin{equation}
 \bar{x}_f = \frac{\overline{M}^{2}(k_0^2)}{2\beta_{f}} ~~,~~
 \overline{M}^{2}(k_0^2)=M^{2}-x(1-x)k_0^2 ~\nonumber ~
\end{equation}
and $\psi(\bar{x}_f)$ is the digamma function~\cite{arfken}. The usual ($B=0$)
vacuum term, eq.(\ref{ivac3}), will be 
regularized through a 3D non-covariant cutoff.
The expressions given in eqs.(\ref{mpionB0},\ref{therm},\ref{ivac4})  can be used for 
the study of the 
neutral meson properties at finite $T$ e $\mu$.  At the Mott temperature, i. e., when  
$k_0$=$m_{\pi^0}$ = $2M$  a singularity appears in the denominator and
this pole has to be treated in a convenient way, since the pion becomes unstable
and can decay in two quarks.
It is possible to perform analytically
the integration in eq.(\ref{ivac3})  obtaining for its real part the expression:
\begin{eqnarray}
 Re (I_{vac}(k_0^2)) &=& -\frac{1}{8\pi^2} \left[   \ln \frac{\Lambda+\epsilon_\Lambda}{\Lambda} 
 \right.
   \nonumber \\
  &-&\theta(2M-k_0) z_0~
   \tan^{-1}\left(\frac{\Lambda}{\epsilon_{\Lambda} z_0}\right) \nonumber \\
 &-&\left.  \theta(k_0-2M)  
 \frac{z_0}{2}   \left( \ln 
 \frac{\Lambda + \epsilon_\Lambda z_0 } 
  {\Lambda - \epsilon_\Lambda z_0 } \right)
    \right] , \label{evacr}
\end{eqnarray}
with $z_0=\sqrt{|1-4M^2/k_0^2|}$.
Analogously the real part of the magnetic vacuum term given in eq.(\ref{int00})  
can also be calculated yielding:
\begin{eqnarray}
 && Re(I(k_0^2,B))=-\frac{1}{32\pi^2}\sum_{f=u,d} \left[ 
- \int^{1}_{0} dx~ \psi\left( \bar{x}_f +1 \right)  \right.   \nonumber \\
&& + \ln \frac{k_0^2}{2\beta_f} - 2 + \frac{2\beta_f}{k_0^2 z_0} 
\ln \left| \frac{z_0-1}{z_0+1} \right| + \nonumber \\
&& +~\theta(k_0-2M)
\left(\ln \left| \frac{1-z_0^2}{4} \right|  + z_0 \ln \left| \frac{z_0+1}{z_0-1} \right| \right) 
+ \nonumber  \\
&& \left. +~\theta(2M-k_0)
\left(\ln \left( \frac{1+z_0^2}{4} \right)  + 2 z_0 \tan^{-1}\frac{1}{z_0} \right) \right]. 
\end{eqnarray}
Notice that the latter expressions have a branch cut at  $k_0=2M$ and they have to be interpreted as the 
Cauchy principal value when $k_0>2M$~\cite{flor}. These two latter expressions extend our previous results for the 
pion mass calculation~\cite{nosso1} including the regime where $k_0>2M$.

Next, we will return to the main focus of the present work ,i. e., to use the alternative zMFIR scheme
in order to regularize the  
expressions for the polarization integral given in eqs.(\ref{mpion_2},\ref{therm}).
From the eqs.(\ref{intzMFIR},\ref{intB0},\ref{intB}) one obtains:
\begin{equation}
 I(k_0^2,B,T)=I_{vac}(k_0^2)+I(k_0^2,B)+I_{T,\mu}(k_0^2)+I_{T,\mu}(k_0^2,B)~, \label{loop-zeta}
\end{equation}
where
\begin{eqnarray}
 I_{vac}(k_0^2)&=&\sum_{f=u,d}  
\int_M^\infty  dE \frac{E\sqrt{E^2-M^2}}{\pi^2}\nonumber\\
&\times&\frac{1}{4E\left(k_0^2-4E^2\right)} ~, \nonumber \\
I(k_0^2,B) &=& \sum_{f=u,d}  \frac{(2\beta_f)^{1/2}}{2\pi^2}
\int_M^\infty  dE E(q_E) {\cal H}_{1/2}(q_E) \nonumber\\
&\times&\frac{1}{4E(q_E)\left(k_0^2-4E(q_E)^2\right)} ~, \nonumber         \\
 I_{T,\mu}(k_0^2)&=&  - \sum_{f=u,d}  
\int_M^\infty  dE \frac{E\sqrt{E^2-M^2}}{\pi^2}\nonumber\\
&\times&\frac{n(E)+\bar{n}(E)}{4E\left(k_0^2-4E^2\right)} ~, \nonumber    \\
 I_{T,\mu}(k_0^2,B)&=&  - \sum_{f=u,d}  \frac{(2\beta_f)^{1/2}}{2\pi^2}
\int_M^\infty  dE E(q_E) {\cal H}_{1/2}(q_E) \nonumber\\
&\times&\frac{n(E(q_E))+\bar{n}(E(q_E))}{4E\left(k_0^2-4E(q_E)^2\right)} ~. \nonumber
\end{eqnarray}
Therefore, we were able to separate exactly the magnetic and 
non-magnetic components of $I(k_0^2,B,T)$ using the zMFIR regularization scheme.  
The only non finite contribution is the usual $B=0$ vacuum term $I_{vac}(k_0^2)$. Performing the 
change of variables
$p=\sqrt{E^2-M^2}$ in the latter expression for $I_{vac}(k_0^2)$ one easily sees that we reobtain 
exactly the expression given in eq.(\ref{ivac3}) and as discussed above its 
regularized contribution is given by eq.(\ref{evacr}). We perform the same change of variables
in the non-magnetic 
medium term $I_{T,\mu}(k_0^2)$ and it is straightforward to obtain:
\begin{equation}
 I_{T,\mu}(k_0^2) = \frac{1}{8\pi^2} \int_0^\infty dp \frac{p^2}{E(p^2-x_0^2)} ~(n(E)+\bar{n}(E)) ~,
\end{equation}
with $x_0^2=k_0^2/4-M^2$. Here, notice that when $x_0^2$ is greater than zero ($k_0> 2M$) it
is necessary to consider 
this integral as a Cauchy principal value. In order to obtain convenient expressions for 
the numerical calculation 
of the magnetic terms, we perform the change of variables:
\begin{equation}
x^2=q_E=\frac{E^2-M^2}{2\beta_f} \nonumber ~.
\end{equation}
Thus, $dEE=2\beta_f~ xdx$, $E=(2\beta_f)^{1/2}\bar{E}_f $ with $\bar{E}_f=\sqrt{x_f+x^2}$ and $x_f\equiv \frac{M^2}{2\beta_f}$. It is
straightforward to show that:

\begin{eqnarray}
I(k_0^2,B)  &=& - \sum_{f=u,d}  \frac{(2\beta_f)^{1/2}}{32\pi^2}
\int_0^\infty  dx x {\cal H}_{1/2}(x^2) \nonumber\\
&\times&\frac{1}{{\bar E}_f\left(x^2-\bar{x}_0^2\right)} ~, \\
 I_{T,\mu}(k_0^2,B)  &=&  \sum_{f=u,d}  \frac{(2\beta_f)^{1/2}}{32\pi^2}
\int_0^\infty  dx x {\cal H}_{1/2}(x^2) \nonumber \\
&\times&\frac{n(E)+\bar{n}(E)}{{\bar E}_f\left(x^2-\bar{x}_0^2\right)}  ~, 
\end{eqnarray} 
\noindent where $\bar{x}_0^2=(k_0^2/4-M^2)/(2\beta_f)$. Of course, the latter expressions also have to be interpreted 
as Cauchy principal values when $\bar{x}_0^2$ is greater than zero or, equivalently, when $k_0> 2M$, i. e., 
the Mott temperature is univocally defined when the system is immersed in a magnetic medium.
The $\pi_0$ pole mass is calculated numerically using the eqs.(\ref{mpionB0},\ref{loop-zeta})
in terms of the Hurwitz-Riemann
zeta functions as discussed above and our results will be presented in the next section. 
%
%
\section{Numerical Results}
\label{numresult}
In the following we present the numerical results. The set of parameters utilized for 
the $3$D cutoff regularization are $\Lambda=664.3$ MeV, $m_0=5.0$ MeV and 
$G=\frac{2.06}{\Lambda^2}$~\cite{buballa}. Firstly, we sketch in Fig.~\ref{fig2} the effective 
quarks masses as well as the pole-mass of neutral meson $\pi_0$ at finite temperature and 
zero magnetic field. At low temperatures the effective quark mass
is almost the same as
calculated in the vacuum, i.e, $M(T)\approx M(0)$, as well as the pole-mass of the $\pi^0$ meson. 
When the pseudo-critical
temperature is exceeded, the  chiral symmetry is partially restored 
and the effective quark mass
becomes almost the current quark masses $M\approx m_0$. 
The Mott dissociation
 happens when the system gets $m_{\pi}(T)=2M(T)$~\cite{blasche} at $T_{Mott}=179.8$ MeV.  
 In the Wigner-Weyl phase, the equations (\ref{IqnB0}) should be interpreted as its Principal 
 Value, due to the divergences in the denominator when 
 $p=\sqrt{\frac{m_{\pi^0}^2}{4}-M^2}$~\cite{flor,asakawa}.

 \begin{figure}[h]
\begin{tabular}{ccc}
\end{tabular}
\end{figure}
\begin{figure}[h]
\begin{tabular}{ccc}
\includegraphics[width=9.cm]{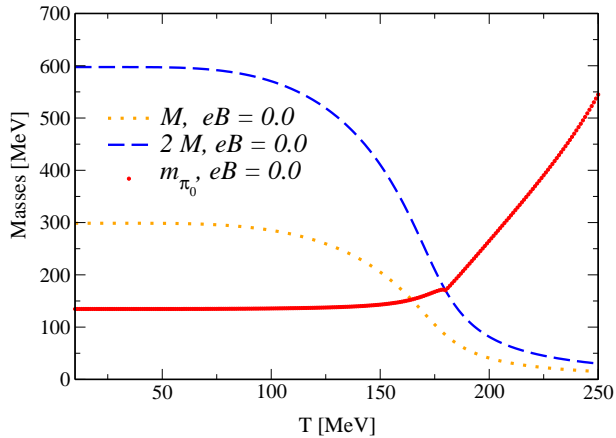}\\
\end{tabular}
\caption{Masses as a function of the temperature for eB=0. }
\label{fig2}
\end{figure}
 
The behavior of the effective quark masses at finite temperature at $eB=0.1$GeV$^2$ can be seen 
in panel (a) in the Fig.~\ref{fig3}. As in the previous case, the pole-mass of 
the neutral collective excitation
is showed in the same figure as well. For the effective quark masses evaluated in both formalisms 
in the present paper, i.e., eq(\ref{gap_eq}) and eq(\ref{gapzmfir}),
almost no difference 
can be seen in the numerical
results as we already mentioned and the  well known behavior is sketched. At low 
temperatures, the magnetic field enhances the breaking of the  chiral symmetry and
the effective quark masses becomes more stronger (magnetic catalysis~\cite{revigor}). When the 
temperature is greater than the pseudo-critical temperature, the chiral symmetry partially 
restores, and the 
effective quark masses becomes weaker. For completeness, we mention that in this scenario, it 
is well established that the pseudo-critical temperature $T_c$ increases with $eB$~\cite{farias,farias2}.

For the pole-masse of the neutral collective excitation, the analysis becomes more complicated. 
For low temperatures($T<T_c$) the behavior of the neutral meson is almost as 
expected, the $m_{\pi^0}(B,T)\approx m_{\pi^0}(B)$. We can understand this in terms of the 
chiral symmetry restoration. In this phase, these mesons are
protected by the Goldstone phase, and the magnetic field is strengthening the break of the partially 
chiral symmetry. 
When the temperature is sufficiently to partially restore the chiral symmetry, the neutral
mesons enters in the Wigner-Weyl phase. In this phase, the $\pi_0$ meson is a thermal
excitation with a finite decay width, but the increase of the magnetic field causes the 
thermal excitations to become more energetic when compared with the zero magnetic field case. 
The $\pi^0$ mass suddenly jumps to a more energetic solution at the dissociation
temperature.
Following the previous analysis, for the $eB=0.1$GeV$^2$, we 
found $T_{Mott}=195$ MeV. 

\begin{figure}[h]
\begin{tabular}{ccc}
\end{tabular}
\end{figure}
\begin{figure}[h]
\begin{tabular}{ccc}
\includegraphics[width=9.cm]{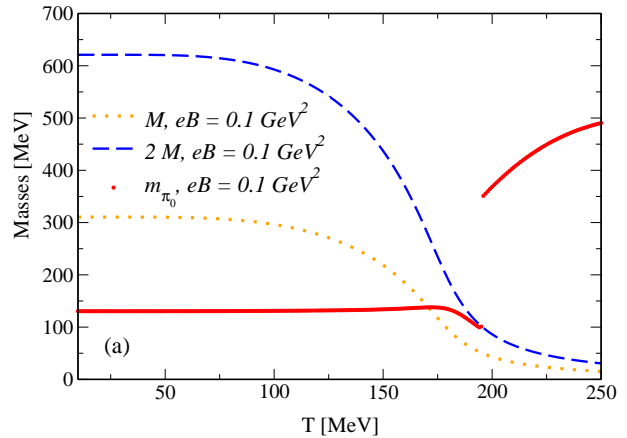}\\
\includegraphics[width=9.cm]{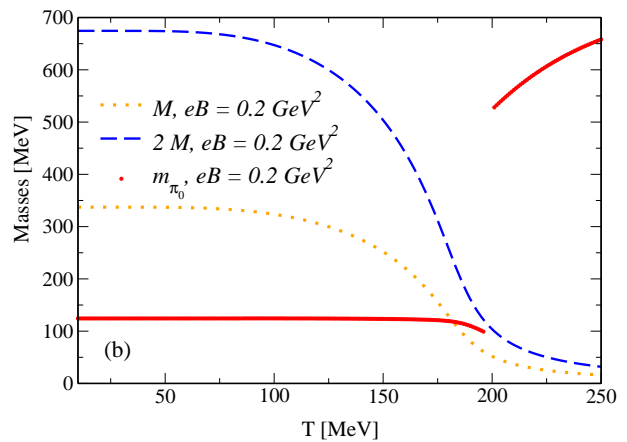}\\
\end{tabular}
\caption{Masses as a function of the temperature for different values of magnetic field.}
\label{fig3}
\end{figure}

In panel (b) of Fig \ref{fig3}  we sketch the results with
$eB=0.2\, GeV^2$, and as expected, at low temperatures the previous analysis can still be 
used, where we have $m_{\pi^0}(B,T)\approx m_{\pi^0}(B)$. 
The interpretation slightly changes if we look beyond the Mott Temperature $T_{Mott}=200.2$ MeV, 
where
the magnetic field enhances the resonant excitations to more energetic states. We show that
using zMFIR  regularization scheme that the effect of the magnetic field is catalyze the Mott 
temperature, result that is
the opposite found in recent literature~\cite{mao}. Other point in Ref.~\cite{mao} that is
not clear is the position of the jump on the pion mass, the authors claim that the jump happens
in the Mott temperature, but we do not observe this in
Fig.2 of Ref.~\cite{mao}, e.g. the jump for eB $= 10\,m_{\pi}^2$ happens when $m_{\pi} = M$, 
and the definition for Mott temperature is $m_{\pi} = 2 M$.

\subsection{The nature of the sudden jump on the pion mass}

In~\cite{mao} the author use Pauli-Villars regularization in NJL model to study the behavior
of the 
neutral pion mass as a function of the temperature for different values of magnetic fields. 
The author presents a good explanation for the sudden jump on the pion mass at the Mott
temperature, he argued that these more energetic thermal excitations appear due the 
divergence at the Lowest Landau Level (LLL) that is associated with the dimensional 
reduction caused by the strong magnetic fields. 

The dimensional reduction of fermionic systems at strong magnetic field has been 
explored by several authors~\cite{gus,fukushima2} and plays a fundamental role to explain
these more energetic resonances. In the MFIR formalism we can clearly see in a more transparent
way that we have a physical dimensional reduction of our system when we compare 
with the $eB=0$ equations. The LLL dominance at strong magnetic field makes our system
go in a naively way to $3+1$ D$ \,\, \rightarrow \,\,1+1$ D, e.g., see eq.(\ref{potmag}) 
when $n\rightarrow 0$ in comparison with eq.(\ref{Evac}). But, if our system is going 
to the high temperature limit, the highest Landau Levels become populated again, 
because of
the weakening of the interaction of the fermions in the chiral condensate.

For the collective excitations, the idea becomes the same. The high temperature($T>T_{Mott}$)
is not sufficient to excite all possible states in the phase space, as we have in the $eB=0$ case, 
that allow the resonant pair $q-\bar{q}$ to occupy. Due to
the dimensional reduction associated with the magnetic field, the number of states 
for the excited pair is reduced, 
and the momentum integration is not sufficiently to guarantee
some solution to the resonant state of the $\pi^0$ at energy states right after $m_{\pi^0}\approx2M$. 
That is, the dimensional reduction
of the system enforces the system to have less states than the $eB=0$ case. On the other hand,
 the condition $1-2G\Pi_{ps}(k_0^2=m_{\pi^0}^2)=0$ is the RPA equation for the pole-mass, that
ensures to us that the conservation of the external momenta in the RPA bubbles should 
be on shell. Therefore, we can expect that the $\pi^0$ mass jump to another
more energetic state at the Mott Temperature. This jump is just the $\pi^0$ mass going
to its lowest possible energy state, when all other states are not accessible anymore.

At the Wigner-Weyl phase, 
we should mention an interesting phenomenon. If we look at the top of figure \ref{fig3}, 
we can see a decrease of 
the energy of the resonant $\overline{q}-q$ state (or the $\pi_0$ mass)
at finite magnetic field $eB=0.1$GeV $^2$ and high 
temperature(say $T=250$ MeV) 
when compared to the same situation at $eB = 0$ of the figure \ref{fig2}. 
On the other hand, if we increase  the magnetic field 
(in our case we see this for $eB=0.2$ GeV$^2$ at the 
bottom of the figure \ref{fig3}), the energy of the resonant
state increases above the $eB=0$ situation and this effect can be understood as an oscillatory behavior. 
The main reason for this to happen is that at $eB=0$ 
most of the energy of the neutral 
resonance is generated thermally. At finite magnetic field the dimensional reduction
that takes place softens the thermal energy.
 At a certain small magnetic field the dimensional reduction and the magnitude of the magnetic field are not 
 sufficiently strong to increase the energy of the resonant pair above the energy associated with $eB=0$ case. 
 If we increase the
magnetic field to $eB = 0.2$ GeV$^2$(as we see in the fig.\ref{fig3}), we will have more energy from 
the magnetic field contributing
to the resonant $\overline{q}-q$  pair overcoming at a certain point the weakening due to the dimensional
reduction, increasing for this reason the energy of the state.

\section{Effects due inverse magnetic catalysis}
\label{imc}

Inverse magnetic catalysis is a remarkable phenomena that was discovered by lattice QCD 
simulations~\cite{bali,lattice}. Recently was shown that IMC can be reproduced in 
Nambu--Jona-Lasinio model if thermo-magnetic effects are include in the coupling 
constant $G(eB,T)$~\cite{farias,farias2}

In order to see how the inverse magnetic catalysis can influence our results, we 
improve the calculations using the same $G(eB,T)$ fitted to the lattice data for the 
average of the quark condensates~\cite{bali}. In~\cite{farias2} this fit was obtained by 
using the following interpolation formula for $G(eB,T)$:
\begin{eqnarray}
G(B,T) = c(B)\left[1-\frac{1}{1+ e^{\beta(B) \left[T_a(B) - T\right]}}\right]+s(B) .
\label{ourGBT}
\end{eqnarray}
The values of the parameters $c,s,\beta$ and $T_a$ are shown in 
Table~\ref{glatnjl}\footnote{Note that the parameters $c, s,\beta$ and $T_a$ depend only on
the magnetic field} and for the 
the parametrization of the model we take standard values, $\Lambda = 0.650~{\rm GeV}$ 
and $m = 5.5~{\rm MeV}$.  
\begin{table}
\caption{Values of the fitting parameters in Eq.~(\ref{ourGBT}). Units are in appropriate 
powers of GeV.}
\label{glatnjl}
\begin{tabular}{|c|c|c|c|c|}
\hline\noalign{\smallskip}
$eB$ & $c$ &  $T_a$ & $s$   & $\beta$  \\ \hline\noalign{\smallskip}
0.0    &    0.900 &  0.168   &  3.731   &  40.000 \\\hline\noalign{\smallskip}
0.2    &    1.226  &  0.168   &  3.262   &  34.117 \\\hline\noalign{\smallskip}
0.4    &    1.769  &  0.169   &  2.294   &  22.988 \\\hline\noalign{\smallskip}
0.6    &    0.741  &  0.156   &  2.864   &  14.401 \\\hline\noalign{\smallskip}
0.8    &    1.289  &  0.158   &  1.804   &  11.506 \\\noalign{\smallskip}
\hline
\end{tabular}
\end{table}
In the Fig.~\ref{fig4} we show the effect of the inverse magnetic catalysis in the effective quark 
masses as well as the pole-mass of the neutral
meson. The Mott temperature
in this case suffers a inverse magnetic catalysis in the same way, passing by $T_{Mott}=166.84$ MeV 
at $eB=0.1$ GeV$^2$ to $T_{Mott}=164.9$  MeV at $eB=0.2$ GeV$^2$.
The resonances as we can see are still more energetic as we grow the magnetic field.
\begin{figure}[h]
\begin{tabular}{ccc}
\end{tabular}
\end{figure}
\begin{figure}[h]
\begin{tabular}{ccc}
\includegraphics[width=8.cm]{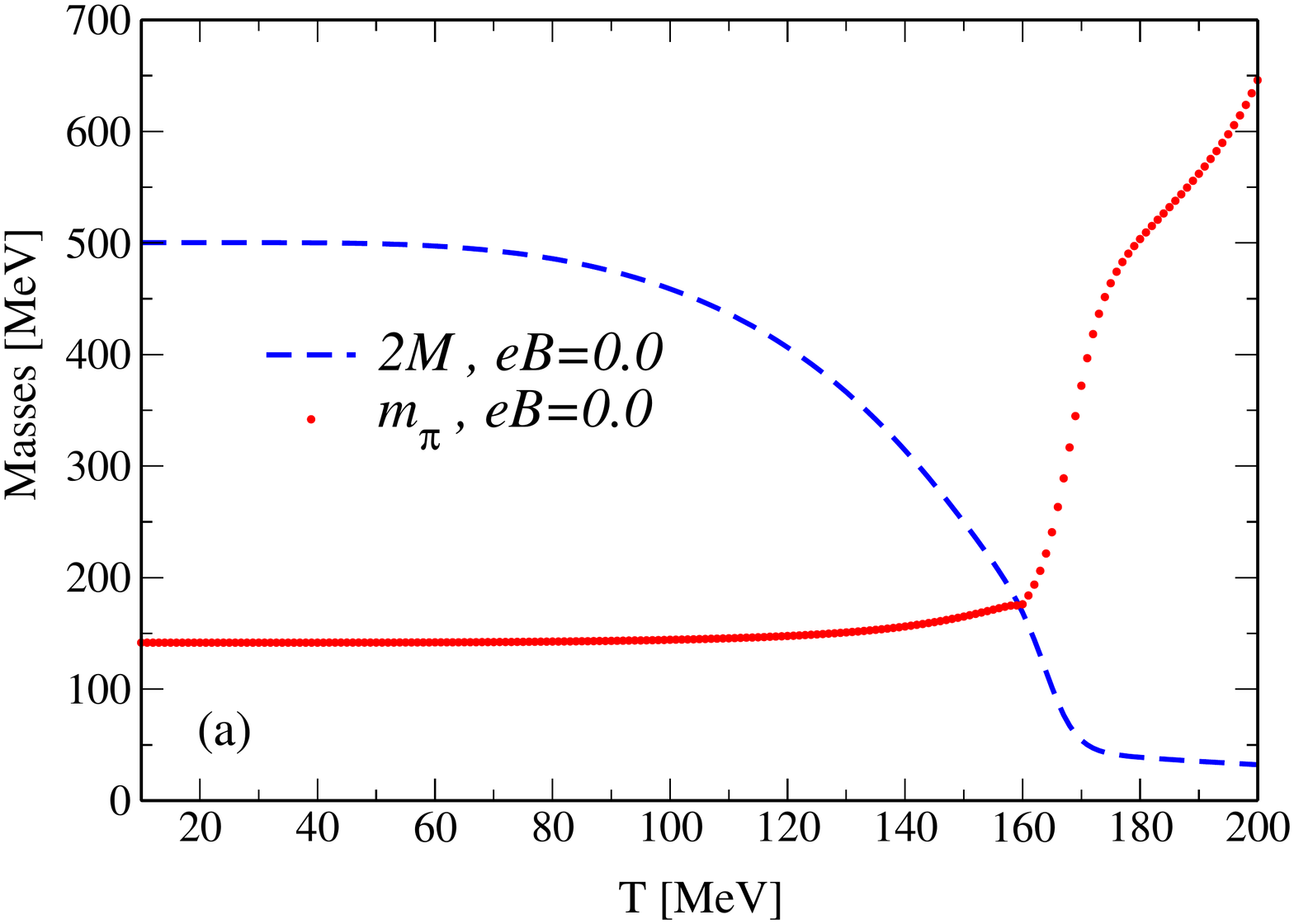}\\
\includegraphics[width=8.cm]{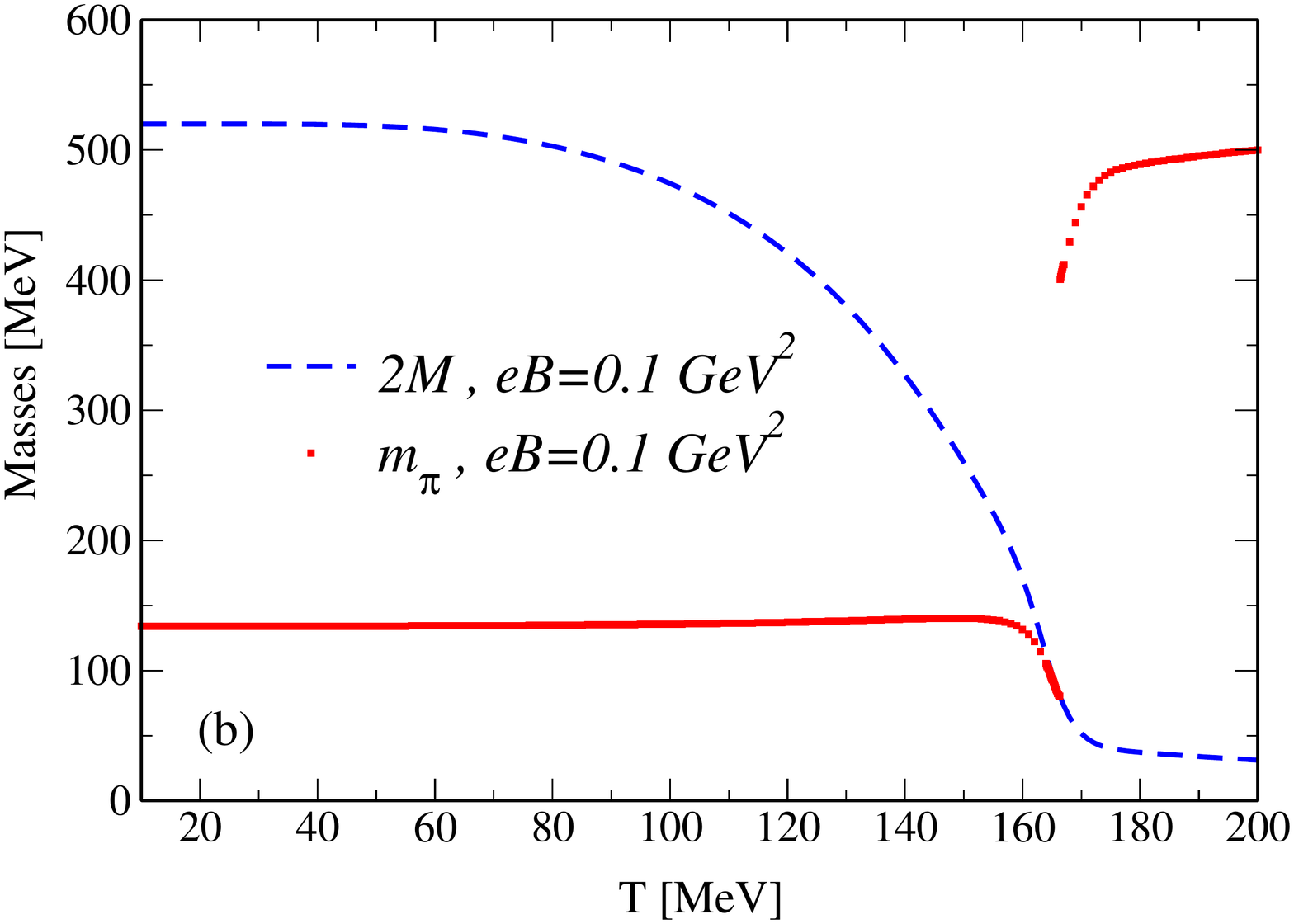}\\
\includegraphics[width=8.cm]{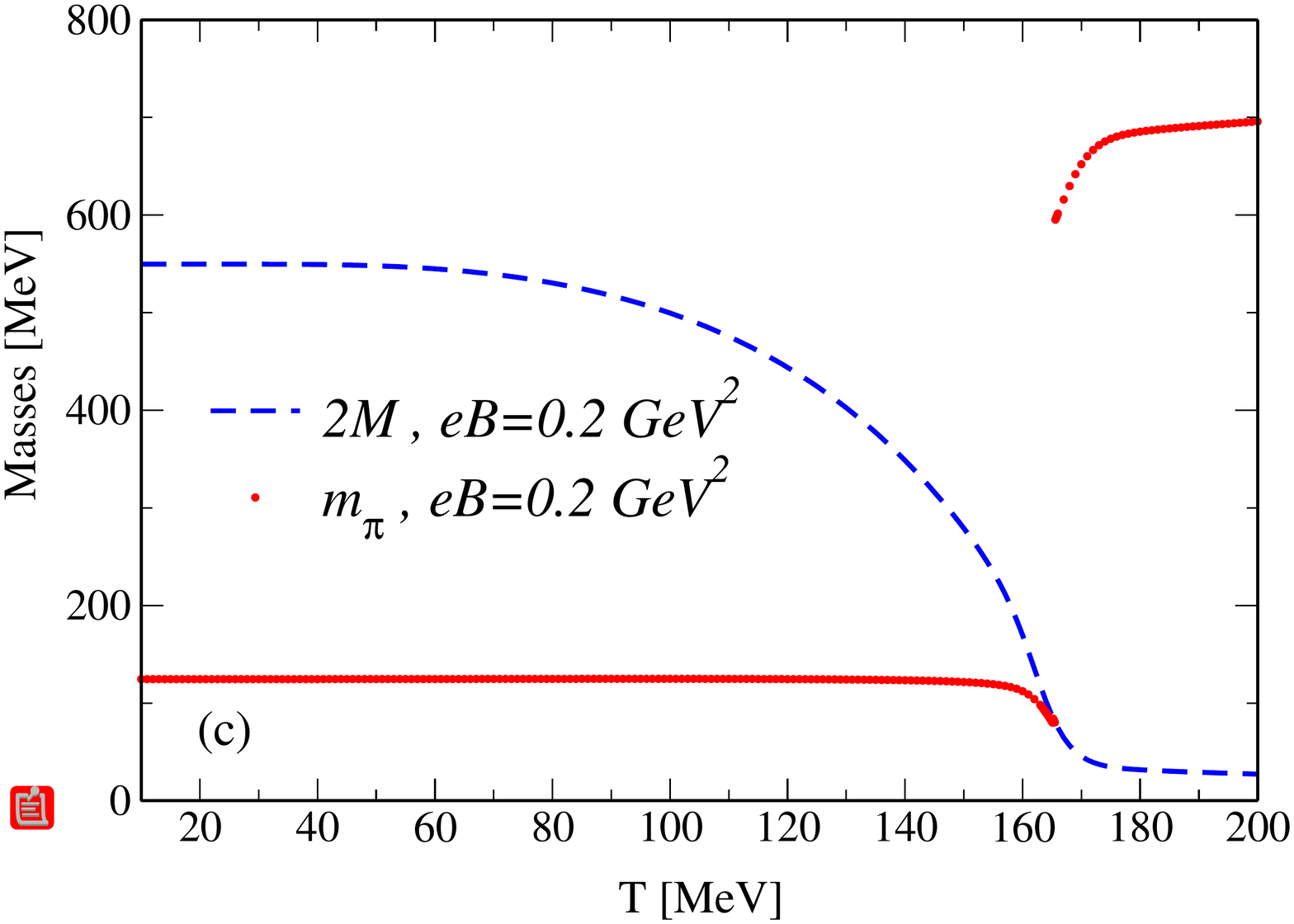}\\
\end{tabular}
\caption{Masses as a function of the temperature for different values of magnetic field using 
the thermo-magnetic dependence on the coupling constant $G(eB,T)$.}
\label{fig4}
\end{figure}  
 
The physical analysis involving the chiral symmetry restoration at strong magnetic fields is the same 
as the one discussed in the previous section. The magnetic field
enhances the binding of the condensate and the broken chiral symmetry is even more evident. As we increase the 
temperature the weakening of the 
chiral condensate happens and the chiral symmetry can be partially restored at high temperatures. 
At this point,
the difference from our previous section is that the coupling constant $G$ is substituted
by a thermo-magnetic dependent coupling $G\rightarrow G(eB,T)$ 
that mimics the inverse magnetic catalysis in our results for the effective quark masses. For the neutral meson mass 
the broken chiral symmetry preserves at low temperature $m_{\pi^0}(B,T) \approx m_{\pi^0}(B,0)$ the pole mass approximately 
as a pseudo-Goldstone boson. 
If we increase the temperature with the coupling $G(eB,T)$ our results show that the Mott temperature decreases compared to
the constant coupling $G$ case, in agreement with what is expected by 
the inverse magnetic catalysis. The oscillatory behavior in the Wigner-Weyl phase, illustrated in the last section, 
also occurs in the present case as we can see from the results of the figure \ref{fig4}.
The physical analysis of this behavior is the same as pointed out in the last section, 
since we have just modified the field dependent coupling $G(eB,T)$. 
 
However, we should pay
attention to the fact that the physical mechanism behind the inverse magnetic catalysis (IMC)
is still under debate, for more discussions~\cite{revigor}.  
 
Also, we have to mention that, some works evaluating the neutral mesons masses at finite magnetic 
fields and temperatures have not obtaining any jump 
in the $\pi^0$ pole-mass at the
dissociation temperature, and this
can cause some misunderstanding. As explained in~\cite{zhuang,iran}, there are some approximations 
in the formalism, an special is the integral $I(k_0^2)\approx I(0)$. 
Another feature that can be explored is
the role of the regularizations that are used and are not fully exploited, since the choice of 
some regularizations can produce significant non-physical results~\cite{norberto,ricardo,scoccola},
differently of the case
of the regularization adopted in this work.
%
%

\section{Conclusions}
\label{conclusions}

In this work we have presented a study of the NJL SU(2) at finite temperatures and magnetic fields 
in the MFIR scheme. In an analogous way, we show an equivalent
and helpful formalism that we call zMFIR, applicable in future several applications.

The new formalism presented in this work has the advantage of separating in 
an exact way the magnetic from non-magnetic contributions for the expressions of physical 
interest, i. e., the thermodynamic potential, gap equation, polarization loops, etc. 
As an example 
of a direct application of the zMFIR, we expand the formalism for the RPA framework, and we evaluate 
the
pole-mass of $\pi^0$ meson at finite magnetic field and temperatures. Again, both formalisms agree 
almost exactly in their numerical results, though
the zMFIR showed to be more efficient to work in the numerical evaluations.
The first dramatic result is the more energetic resonances that we obtain as we increase the magnetic 
field.  As we explained, this is a direct result from the dimensional reduction of 
the system at strong magnetic
fields that enforces the system to go to another state, since we have less states to the creation 
of the thermal $q-\bar{q}$ excitation.

As mentioned in~\cite{norberto,ricardo,scoccola}, the MFIR scheme avoid some unphysical results, and
this choice of regularization provide to us some different results from most
of the regularizations prescriptions of
the current literature. The Mott dissociation temperature is catalyzed with the increase of the 
magnetic field as well as the pseudo-critical temperature.
This behavior of
the Mott temperature goes in a inverse way of the reference~\cite{mao}. 
We believe that some differences
associated with the regularization adopted in that work can be affecting
the way the mesons are becoming resonances. Also, the fact that we are working in a mean field
approximation strengthens the idea that in this approximation, we 
have to see the catalyzed Mott Temperature. 

To recover the expected result from Lattice~\cite{bali} we also employed the $G(eB,T)$ 
from~\cite{farias2}. The inverse magnetic catalysis in the NJL SU(2) model 
as discussed in
the main text
is recovered, 
and in the same way we obtain a inverse magnetic catalysis of the Mott Temperature. 

As a suplementary result, the magnetic vacuum polarization integral of 
reference\cite{nosso1} has been extended by using the analytic continuation technique. 
These analytical expressions are useful for the 
validation of more complicated numerical calculations.

\section*{Acknowledgments}

We thank Norberto Scoccola for useful discussions. This work was  partially 
supported by Conselho Nacional de Desenvolvimento 
Cient\'{\i}fico e Tecnol\'{o}gico (CNPq) under grants 304758/2017-5 (R.L.S.F) 
and 6484/2016-1 (S.S.A.), 
and as a part of the project INCT-FNA (Instituto Nacional de Ci\^encia e Tecnologia -
F\'{\i}sica Nuclear e Aplica\c c\~oes) 464898/2014-5 (SSA)
and Coordena\c c\~ao de Aperfei\c coamento de Pessoal de N\'ivel Superior (CAPES) (W.R.T).
%
\appendix
\section{regularization of the magnetic effective potential term} 
\label{appA}

Next, we will discuss how to identify and separate the divergent
contribution to the  vacuum magnetic 
effective potential term $\tilde{\Omega}_{vac}(B)$.
This latter term can be obtained  by substituting 
the explicit form of $\tilde{g}_f(q_E,B)$, eq.(\ref{gEZETA}), in eq.(\ref{EvacB}):
\small{
 \begin{eqnarray}
  &&\tilde{\Omega}_{vac}(B)=-N_c\sum_{f=u,d} \frac{(2\beta_f)^{2}}{(2\pi)^2}\int_0^\infty dq_E
                                                \sqrt{x_f+q_E}    \nonumber \\ 
  && \times \left[ \zeta(\frac{1}{2},\{q_E\}) - 
  \left( \zeta(\frac{1}{2},q_E) +2 q_E^{1/2}-\frac{1}{2q_E^{1/2}} \right) \right] ~,                                                                                                                                                                                               
  \label{EvacBR}
 \end{eqnarray}  }
 
\noindent where $x_f=\frac{M_f^2}{2\beta_f}$. 
From the asymptotic limit of the
zeta function $\zeta(1/2,q_E)$~\cite{DE},
\begin{equation}
\zeta(\frac{1}{2},q_E)=-2q_E^{1/2}+\frac{1}{2q_E^{1/2}}+
\frac{1}{24}q_E^{-3/2}+O(q_E^{-7/2})~~,
\end{equation}
we see that the integrand in eq.(\ref{EvacBR}) is logarithmically divergent. Hence, we will sum and 
subtract the term
$-\frac{1}{24}q_E^{-3/2}$ in the integrand, since this term forces the convergence of the integral.
Therefore, one obtains: 
\small{
 \begin{eqnarray}
&& \tilde{\Omega}_{vac}(B)=-N_c\sum_{f=u,d} \frac{(2\beta_f)^{2}}{(2\pi)^2} \left[ \frac{}{} \right. 
                                                                                        \nonumber \\
 &&  -\int_0^\infty dq_E \frac{\sqrt{x_f+q_E}}{24 q_E^{3/2}}  +
 \int_0^\infty dq_E  \zeta(\frac{1}{2},\{q_E\}) \sqrt{x_f+q_E} \nonumber \\ 
&& \left. -\int_0^\infty dq_E \left(  
  \zeta(\frac{1}{2},q_E) +2 q_E^{1/2}-\frac{1}{2q_E^{1/2}} -\frac{1}{24q_E^{3/2}}\right) 
                                                           \sqrt{x_f+q_E} \right]. \nonumber \\ 
                                                           \label{OMvac5}
 \end{eqnarray} }
The first integration in the latter equation is trivial :
\begin{equation}
 \int_0^\infty dq_E \frac{\sqrt{x_f+q_E}}{q_E^{3/2}} =
\left. -2 \left(\frac{\sqrt{x_f+q_E}}{q_E^{1/2}} + 
 \ln [ q_E^{1/2}+\sqrt{x_f+q_E}]\right) \right|_0^\infty ~. \label{inttriv}
\end{equation}
The integral involving the periodic function, $\zeta(\frac{1}{2},\{q_E\})$,  in 
eq.(\ref{OMvac5}) can be 
rewritten by noting that:
\begin{eqnarray}
&& \int_0^{\infty}dx f(x) g(x) = 
 \int_0^{1}dx f(x) g(x) + \int_1^{2}dx f(x) g(x)  \nonumber \\ 
&& + \int_2^{3}dx f(x ) g(x) + \dots  =\int_0^{1} dx f(x) \sum_{k=0}^\infty  g(x+k)   ~ , 
\label{oscint}
\end{eqnarray}
where f(x) is a periodic function with period 1 and  g(x) an arbitrary function. Notice that
a change of variables was done to set the limits of integration of all integrals 
from 0 to 1. Then, it follows that: 
\begin{eqnarray}
 &&\int_0^{\infty}dq_E  \zeta(\frac{1}{2},\{q_E\})  \sqrt{x_f+q_E} = \nonumber \\
 && \int_0^{1}dq_E  \zeta(\frac{1}{2},q_E) \sum_{k=0}^\infty  \sqrt{x_f+q_E +k}~.
 \label{change01}
\end{eqnarray}
From the definition of the Hurwitz-Riemann zeta function, eq.(\ref{HZ}), the latter integral
may be written as:
\begin{eqnarray}
&& \int_0^{1}dq_E  \zeta(\frac{1}{2},\{q_E\}) \sum_{k=0}^\infty \sqrt{x_f+q_E +k}~                                                                                   \nonumber \\
&&= \int_0^{1}dq_E ~  \zeta(\frac{1}{2},q_E)~  \zeta(-\frac{1}{2},x_f+q_E) ~.\label{intnontriv}
\end{eqnarray}
\noindent
At this point, we substitute eqs.(\ref{inttriv},\ref{intnontriv}) in eq.(\ref{OMvac5}).
\noindent Finally, aiming a more suitable expression for numerical calculations, we use the 
following property of the Riemann-Hurwitz zeta function~\cite{apostol}
\begin{equation}
 \frac{\partial}{\partial y} \zeta(z,y)=-z\zeta(z+1,y)~, \label{derzeta}
\end{equation}
 in order to perform
an integration by parts in eq.(\ref{OMvac5}).
 Then, the final regularized finite magnetic vacuum term is given by:
 \begin{eqnarray}
\tilde{\Omega}_{vac}(B)&=&-N_c\sum_{f=u,d} \frac{(2\beta_f)^{2}}{(2\pi)^2}~ \left\{ \frac{2}{24} +
                                                 \frac{1}{24} \ln x_f  \right. \nonumber \\ 
&-&\int_0^{1}dq_E    \zeta(-\frac{1}{2},\{q_E\}) \zeta(\frac{1}{2},q_E+x_f)  
                                                     \nonumber \\
&+&  \int_0^{\infty}dq_E \left[  \zeta(-\frac{1}{2},q_E) 
+ \frac{2}{3}q_E^{3/2}  \right. \nonumber \\
&-& \left. \left. \frac{1}{2}q_E^{1/2} + \frac{1}{24}q_E^{-1/2} \right]
                                              \frac{1}{\sqrt{x_f+q_E}}   \right\}  ~.   
 \end{eqnarray}

This integral is adequate for numerical calculations and we have discarded 
all the unphysical divergent terms. As discussed in section II.A, 
the latter regularized magnetic 
vacuum expression  numerically coincides with the one obtained by using
other formalisms.
%
%
%
%
\section{the equivalence between MFIR and zMFIR} 
\label{appB}

Here, we will present the main steps for proving the equivalence between   
the zMFIR and MFIR formalisms for obtaining the gap equation.

Starting from eq.(\ref{gap_zmfir}) at $T=0$:

\begin{equation}
 \frac{M-m}{2MG}=~I_G+I_G(B)~,\label{gap_zmfir0}
\end{equation}

\noindent where $I_G$ is the vacuum contribution and $I_G(B)$ is given by:

\begin{eqnarray}
I_G(B)=N_c\sum_{f=u,d} \int_{0}^{\infty} ~dq_E~~\frac{(2\beta_f)^{3/2}}{(2\pi)^2} 
                      \frac{\tilde{\cal H}_{1/2}(q_E)}{E(q_E)} ~ \label{gapzmfir0}
\end{eqnarray}

\noindent and $E(q_E)$ and $\tilde{\cal H}_{1/2}(q_E) $ are 
given in eqs.(\ref{EqE},\ref{gEZETA1}) respectively.
\noindent We can split eq.(\ref{gapzmfir0}) in two contributions:

\begin{eqnarray}
&&I_G(B)=N_c\sum_{f=u,d} \left[I(x_f)_{z}+I(x_f)_{div}\right], \label{gapzmfir1}  
\end{eqnarray}

\noindent in the last equation:

\small
\begin{eqnarray}
I(x_f)_z&=& \frac{(2\beta_f)}{(2\pi)^2} \int_{0}^{\infty} ~dq_E~~ 
                      \frac{\left(\zeta(\frac{1}{2},\{q_E\})  - 
                      \zeta(\frac{1}{2},q_E)\right)}{(x_f+q_E)^{1/2}} ,\label{Iz01}\\
I(x_f)_{div}&=&\frac{(2\beta_f)}{(2\pi)^2}  \int_{0}^{\infty} ~dq_E~~  
                      \frac{\left(-2q_E^{1/2}+\frac{1}{2q_E^{1/2}}\right) }{(x_f+q_E)^{1/2}}
                      .\label{Idiv01}
\end{eqnarray}

These two integrals can be evaluated by using an integration by parts  and the 
eq.(\ref{derzeta}), yielding 

\small
\begin{eqnarray}
I(x_f)_z&=&\frac{(2\beta_f)}{(2\pi)^2}  \int_{0}^{\infty} dq_E
                      \frac{\left(\zeta(-\frac{1}{2},\{q_E\})  - \zeta(-\frac{1}{2},q_E)\right)}{(x_f+q_E)^{3/2}} ,\label{Iz02}\\
I(x_f)_{div}&=& \frac{(2\beta_f)}{(2\pi)^2}  \int_{0}^{\infty} dq_E  
                      \frac{\left(-\frac{2}{3}q_E^{3/2}+\frac{q_E^{1/2}}{2}\right) }{(x_f+q_E)^{3/2}} .\label{Idiv02}
\end{eqnarray}

Next, we use the same procedure of Appendix~\ref{appA} in order to rewrite 
the first of these two integrals. First, we use eq.(\ref{oscint}) to obtain:   
\begin{eqnarray}
 &&\int_0^{\infty}dq_E  \zeta(-\frac{1}{2},\{q_E\})  \frac{1}{(x_f+q_E)^{3/2}} = \nonumber \\
 && \int_0^{1}dq_E  \zeta(-\frac{1}{2},q_E) \sum_{k=0}^\infty  \frac{1}{(x_f+q_E)^{3/2}}~.
\end{eqnarray}
From the definition of the Hurwitz-Riemann zeta function, eq.(\ref{HZ}), 
the last sum of integrals
may be written as:
\begin{eqnarray}
&& \int_0^{1}dq_E  \zeta(-\frac{1}{2},\{q_E\}) \sum_{k=0}^\infty  \frac{1}{(x_f+q_E)^{3/2}}~                                                                                   \nonumber \\
&&= \int_0^{1}dq_E ~  \zeta(-\frac{1}{2},q_E)~  \zeta(\frac{3}{2},x_f+q_E) ~.\label{change02}
\end{eqnarray}
Returning to eq.(\ref{Iz02}), we can rewrite $I(x_f)_z$:
\small
\begin{eqnarray}
I(x_f)_{z}&=& \frac{(2\beta_f)}{(2\pi)^2}  \left( \int_{0}^{1} ~dq_E~~ 
\zeta(-\frac{1}{2},q_E)\zeta(\frac{3}{2},x_f+q_E) \nonumber \right.\\
&-&\left.  \int_{0}^{\infty} ~dq_E \frac{\zeta(-\frac{1}{2},q_E)}{(x_f+q_E)^{3/2}}\right).\label{Iz03}
\end{eqnarray}

We will now make use of the following  mathematical 
identities(all of them are valid  for $q_E>1$)\cite{apostol2}:
\small
\begin{eqnarray} 
\zeta(\frac{3}{2},x_f+q_E)&=&\frac{1}{\Gamma(3/2)}\int_0^{\infty}dy\frac{y^{1/2}e^{-y (x_f+q_E)}}{1-e^{-y}},\label{zeta01} \\
\zeta(-\frac{1}{2},q_E)&=&\frac{\Gamma(3/2)}{2\pi i}\int_{\mathcal{C}} dz\frac{z^{-3/2}e^{z q_E}}{1-e^{z}},\label{zetaneg02}\\
\frac{1}{(x_f+ q_E)^{3/2}}&=&\frac{1}{\Gamma(3/2)}\int_0^{\infty}dy y^{1/2}e^{-y(q_E+x_f)},\label{id01}
\end{eqnarray}

\noindent where the integral in identity
eq.(\ref{zetaneg02}) is defined  over the Hankel contour $\mathcal{C}$. 
Applying these identities to eq.(\ref{Iz03}), we can perform the integrations 
with respect to $q_E$ getting the result:

\small
\begin{eqnarray}
 I(x_f)_z=\frac{(2\beta_f)}{(2\pi)^2} \frac{1}{2\pi i}\int_{0}^{\infty}dy \frac{y^{-1/2}e^{-yx_f}}{e^{y}-1}\int_{\mathcal{C}}dz\frac{z^{-3/2}}{1-\frac{z}{y}}.\label{Iz04}
\end{eqnarray}

On the contour $\mathcal{C}$ we will use the parametrizations $z=re^{-i\pi}$ 
below the negative 
real axis and $z=re^{i\pi}$ 
above the real axis with $0<r<\infty$. As the first argument of the Riemann-Hurwitz zeta 
function eq.(\ref{zetaneg02}) is negative, 
we can explore its analytic continuation.
With these definitions, one obtains for the contour integration:

\begin{eqnarray}
 \int_{\mathcal{C}}dz\frac{z^{-3/2}}{1-\frac{z}{y}}&&=-\int_{0}^{\infty}dr e^{-i\pi}(e^{-i\pi}r)^{-3/2}(1-\frac{e^{-i\pi}r}{y})^{-1}\nonumber\\
                                                    &&+\int_{0}^{\infty}dr e^{i\pi}(e^{i\pi}r)^{-3/2}(1-\frac{e^{i\pi}z}{y})^{-1},\nonumber\\
                                                   &&=2i\sin(-\frac{\pi}{2} ) y^{-1/2}B(-1/2,3/2), 
                                                   \label{ixf2}
\end{eqnarray}

\noindent where $B(a,b)$ is the Beta function~\cite{apostol} given by
\begin{equation}
B(a,b)=\int_0^{\infty}t^{a-1}(1+t)^{-a-b} ~. \label{betaf}
\end{equation}
where the latter expression can be written in terms of gamma functions as 
$B(a,b)=\frac{\Gamma(a)\Gamma(b)}{\Gamma(a+b)}$.
Afterwards, using  the results of eqs.(\ref{ixf2},\ref{betaf}) in 
eq.(\ref{Iz04}), one obtains:
\begin{eqnarray}
 I(x_f)_z=\frac{(2\beta_f)}{(2\pi)^2}\int_0^{\infty} dy\frac{e^{-yx_f}y^{-1}}{e^{y}-1}.\label{Iz05}
\end{eqnarray}
This quantity diverges at the origin. We can use the series 
expansion $\frac{1}{e^y-1}\approx
\frac{1}{y}-\frac{1}{2}+\mathcal{O}(y)$ to separate the divergent
part of the latter expression. By 
adding and subtracting the first terms of the series in eq.(\ref{Iz05}) one obtains:
\begin{eqnarray}
 I(x_f)_z&=&\frac{(2\beta_f)}{(2\pi)^2}\int_0^{\infty} dy
 \frac{e^{-yx_f}}{y}\left(\frac{1}{e^y-1}-\frac{1}{y}+\frac{1}{2}\right)\nonumber\\
 &+&\frac{(2\beta_f)}{(2\pi)^2}\left(\int_0^{\infty}dy\frac{e^{-yx_f}}{y^2}-
 \int_0^{\infty}dy\frac{e^{-yx_f}}{2y}\right).\label{Iz06}
\end{eqnarray}

The quantity $I(x_f)_{div}$ cancels out the last two integrals in
 eq.(\ref{Iz06}).
This can be seen using the identity, eq.(\ref{id01}), in $I(x_f)_{div}$. 
Noticing that 
$\frac{1}{e^{y}-1}=-\frac{1}{2}+\frac{1}{2}\coth\left(\frac{y}{2}\right)$ and performing the change of
variables $\frac{y}{2\beta_f}=\mu\rightarrow \frac{y}{2\beta_f}=d\mu$, we can rewrite 
eq.(\ref{gapzmfir1}) as:
\begin{eqnarray}
&&I_G(B)=N_c\sum_{f=u,d} \left[I(x_f)_{z}+I(x_f)_{div}\right],\nonumber\\
&&=N_c\sum_{f=u,d}\frac{\beta_f}{(2\pi)^2}\int_0^{\infty} 
d\mu\frac{e^{-\mu M_f^2}}{\mu^2}
\left(\mu\coth\left(\beta_f \mu\right)-\frac{1}{\beta_f}\right).\nonumber\\
&&=N_c\sum_{f=u,d}\frac{\beta_f}{2\pi^2}\left[\log\Gamma(x_f)-(x_f-\frac{1}{2})\log(2\pi)+ \right.
\nonumber \\
&&\left. x_f -\frac{1}{2}\log 2\pi\right].\nonumber
\end{eqnarray}
If we substitute the last result in eq.(\ref{gap_zmfir0}), 
the usual expression for the gap equation calculated within the MFIR scheme\cite{nosso2}
will be reobtained.
%

\end{document}